# Do 5-minute oscillations of the Sun affect the magnetosphere and lithosphere of the Earth?

A. V. Guglielmi

Institute of Physics of the Earth RAS, Moscow, Russia, *guglielmi@mail.ru*

A. S. Potapov

Institute of Solar-Terrestrial Physics SB RAS, Irkutsk, Russia, *alpot47@mail.ru*

**Abstract.** In this paper, we analyzed a number of problems that are relevant for the physics of solar-terrestrial relationships and the dynamics of geospheres. It is shown that permanent 5-minute oscillations of the photosphere, well known in helioseismology, probably influence regime of the ultra low frequency (ULF) geoelectromagnetic waves. We detected maxima at a period close to 5 min in the spectra of the ULF waves of SE, BMP and Pc5 types. As for the question of the influence of 5-minute oscillations of the photosphere on geoseismicity, the answer is negative. This could be foreseen in advance, but the very formulation of the question led unexpectedly to a turn of the topic and gave us an opportunity to discuss the problem of the hidden periodicity of natural processes synchronized by the world clock. The article was written during the authors' visit to the GO Borok IPE RAS in the summer of 2018.

**Keywords:** gelioseismology, interplanetary plasma, Alfvén waves, Doppler effect, ultra-low-frequency oscillations, earthquakes, thunderstorms, human impact

PACS 94.30.Ms

# 1. Introduction

The impact of solar activity on geophysical processes is a central problem in the physics of solar-terrestrial relations. One of the specific manifestations of solar activity is the permanent oscillations of the photosphere. The question of the geoeffectiveness of the oscillations of the photosphere with a period of 2 h 40 min was posed in the paper [Guglielmi et al., 1977]. In the present paper we will focus on the oscillations of the photosphere with a period of 5 min, well known in helioseismology [Leighton et al., 1962; Ulrich, 1970; Vorontsov, Zharkov, 1981].

The signs of the effect of 5-minute solar oscillations on the oscillations of the Earth's magnetosphere were found in a number of recently published papers [Guglielmi et al., 2015a,b; Guglielmi, Potapov, 2017a; Dovbnya et al., 2017; Dovbnya, Potapov, 2018]. We give an overview of these papers, but first we briefly outline the data on observations of Alfvén waves with periods close to 5 min in the interplanetary medium [Potapov et al., 2012, 2013]. Apparently, the Alfvén waves are excited by the oscillations of the photosphere, and then they are carried by the solar wind to the Earth and affect the mode of oscillations of the magnetosphere.

Getting to work on the review, we also asked ourselves: Do 5-minute oscillations of the Sun affect the seismicity of the Earth? At first glance, the question seems far-fetched. After all, the large distance separating the photosphere from the lithosphere, the massiveness of the lithosphere, which sharply distinguishes it from the magnetosphere, and, most importantly, the uncertainty of the concepts concerning a mechanism of the connection between the photosphere and the lithosphere, seem to make a hopeless an attempt to get a positive answer to the question posed. Nevertheless, guided by the research instinct, we set the task of analyzing global seismicity for the detection of a spectral peak at a period of 5 minutes. The result of the pilot analysis gave us an excuse to discuss in this paper an intriguing problem of anthropogenic impact on the magnetosphere and lithosphere.

## 2. Alfvén waves

The existence of Alfvén waves [Alfvén, 1950] in the interplanetary medium in the Earth's orbit, running from the Sun, is convincingly proved in [Belcher et al., 1969; Belcher, Davis, 1971]. In [Potapov et al., 2012, 2013], the question was raised whether there is a pronounced peak in the spectrum of these waves at a frequency of 3.3 mHz, i.e. at the frequency of the permanent oscillations of the photosphere? A positive answer to this question, firstly, would indirectly confirm the plausible assumption about the transformation of the energy of helioseismic oscillations into the energy of Alfven waves. Secondly, it would testify to the effective propagation of Alfvén waves of this frequency from the Sun to the Earth. Finally, a positive answer to the question posed would give a physical justification for our attempt to look for traces of helioseismic oscillations in the dynamics of the magnetosphere, and possibly also of the Earth's lithosphere.

For the study, ground-based optical observations of the photosphere were used, which were compared with measurements of the oscillations of the interplanetary magnetic field (IMF) from WIND spacecraft near the Earth's orbit. The Doppler shift of the FeI 6569 Å line of solar radiation from 04:47 to 05:47 UT was measured on August 4, 2005 with a 1-s cadence in the region of the coronal hole in the Sun's northern hemisphere. Figure 1 shows the spectra of plasma velocity variations along the line of sight (thin green lines), and the thickened black line shows a spectrum averaged over 127 series of observations along the entire slit of the spectrograph. The maximum of the spectral density corresponds to the interval of oscillation periods between 4.6 and 5.1 minutes.

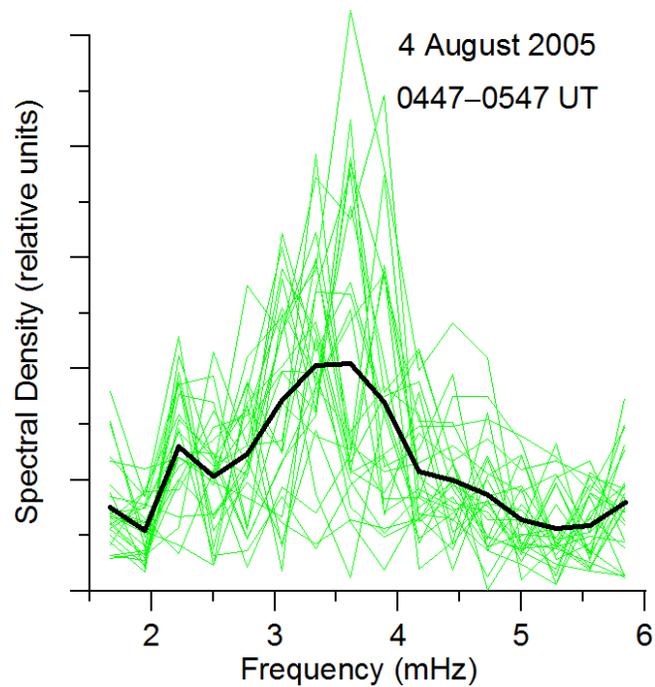

Fig. 1. The spectra of individual time series (thin green lines) of solar oscillations from observations along the slit of the spectrograph and a spectrum averaged over all series (the black line) [Potapov et al., 2012].

It is known that solar plasma is emitted from coronal holes by high-speed solar wind (SW) flows. At the same time, the oscillations observed there could be captured and taken out in the form of Alfvén waves. The idea was to try to catch these waves in the Earth's orbit. The high speed stream from this coronal hole was observed by spacecraft in the Earth's orbit on August 5–7, which roughly corresponded to the time of oscillation transfer by the stream.

In Figure 2, the thickened black line shows the hourly mean values of the SW velocity. The black triangle on the time scale shows the moment of expected arrival of the solar plasma to the Earth's orbit if it was taken out at the time of observation of the oscillations and was transported by a high speed stream with a constant speed of 700 km s$^{-1}$. As we can see, this moment was close to the instant of observation of the maximum SW velocity in the stream. The gray line shows the variations of the average amplitude of ULF waves. It can be seen that the high speed stream brought with it the intensification of ULF oscillations in the solar wind. The main maximum of the amplitude of the ULF waves is ahead of the peak of the SW velocity by approximately 15 hours.

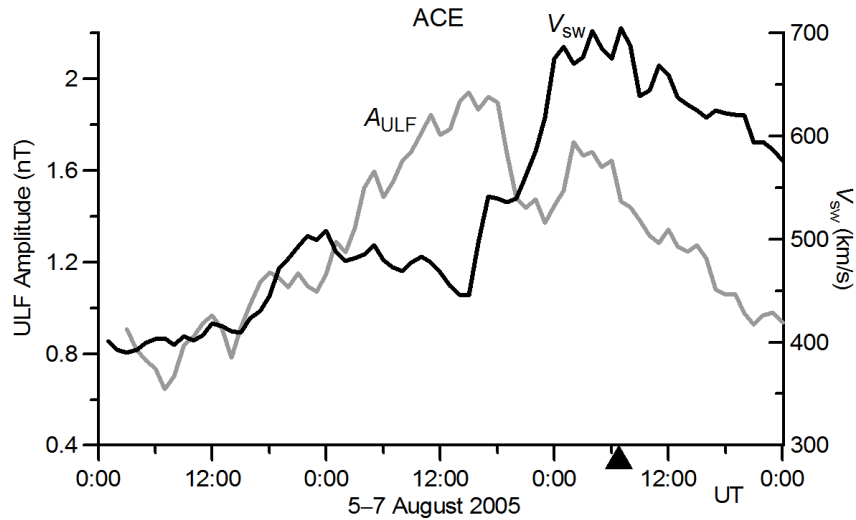

Fig. 2. High speed stream of solar wind from the coronal hole, flowing around the Earth on August 5–7, 2005. The black curve is the hourly mean values of the SW velocity. The gray line is the moving average of the amplitude of ULF oscillations over five hour amplitude values.

The spectrum of ULF oscillations in the solar wind is rather complicated. Clearly, it consists of different sources. It is not easy to find the contribution of solar oscillations. It can be expected that in some SW jets, oscillations of solar origin will make an appreciable contribution to the overall spectrum; in others they will be lost against the background of fluctuations from other sources. For the search, the IMF oscillations were taken from the data of the WIND magnetometer, measured on August 6, 2005 from 05:00 to 19:00 UT, that is, in the region of the maximum amplitude of ULF oscillations. The spectra, constructed on the basis of hourly intervals, showed the presence of peaks in the frequency range from 3 to 4 mHz, an example is shown in Figure 3.

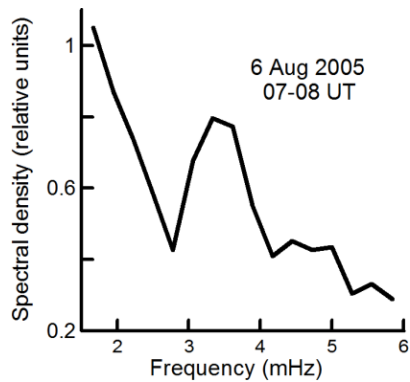

Fig. 3. An example of the spectrum of the hour interval of solar wind oscillations.

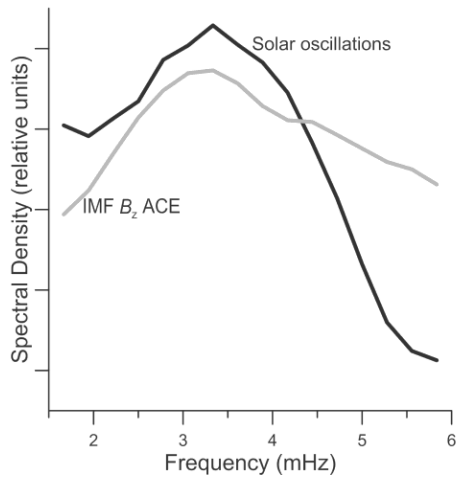

Fig. 4. Comparison of the averaged profiles of the spectral density of oscillations on the Sun (black line) and in the solar wind (gray line).

The spectrum of ULF oscillations of the solar wind averaged over all used hour intervals is compared in Figure 4 with a spectrum of solar oscillations. It is seen that in both cases the peak of the spectral density falls in the interval between 3 and 4 mHz.

The results obtained allow one to affirmatively answer the question posed at the beginning of this section about the presence of a pronounced peak in the frequency spectrum of the Alfvén waves of the solar wind at a frequency that coincides with the dominant frequency of oscillations of the photosphere.

### 3. Oscillations of the magnetosphere

**3.1. Serpentine emission.** In addition to Alfvén waves, ion cyclotron (IC) waves also permanently exist in the solar wind [Guglielmi, 1979; Guglielmi, Pokhotelov, 1996]. Unlike the Alfvén waves excited by the Sun, the IC waves are self-excited in the interplanetary medium as a result of instability. The cause of self-excitation is the velocity anisotropy of the distribution of plasma ions in the solar wind. The anisotropy arises as a result of the adiabatic expansion of the solar corona. It should be recalled that according to theory [Parker, 1963] the solar wind is characterized by permanent ion temperature anisotropy. This fundamental property specifies the fact of a continuous self-excitation of small-scale IC waves in the interplanetary plasma.

We will not reproduce here the rather cumbersome analysis of the dispersion equation of the IC waves (see [Guglielmi, Potapov, 2017a] and the literature cited in this article). Instead, we immediately give an expression for the growth rate of waves of infinitesimally small amplitude in the co-moving reference system:

$$\gamma(k_\parallel, \theta) = \mu \left( k_* - k_\parallel \right) \exp\left[ -\left( k_0/k_\parallel \right)^2 \right] - \eta \theta^2 . \qquad (1)$$

Here $k_\parallel = |k_z|$ in the Cartesian system with $z$-axis along the interplanetary magnetic field **B**, **k** is the wave vector, $\theta$ is the angle between the vectors **B** and **k**, $\theta^2 \ll 1$ (paraxial approximation). We see that the growth rate is dependent on four parameters: $\mu$, $k_*$, $k_0$, и $\eta$. We will not give here the formulas for the relation between these parameters from one hand and the parameters of interplanetary plasma from the other, since they are rather cumbersome. For us, it is important only that $k_* > 0$, $\mu > 0$, $\eta > 0$ in a typical case. Besides, it is important that the increment $\gamma(k_\parallel, \theta)$ has a maximum at $\theta = 0$ and $k_\parallel = k_m \sim \omega_{0p}/c$. Here $\omega_{0p} = \sqrt{4\pi e^2 N/m_p}$, where $e$ and $m_p$ are the charge and mass of the proton, and $N$ is the concentration of electrons in the interplanetary plasma. Note also that $\omega' \sim \Omega_p$ at $k_\parallel = k_m$. Here $\omega'$ is the frequency in the co-moving reference system, $\Omega_p = eB/m_p c$ is the proton gyrofrequency, and $B$ is the modulus of IMF.

Our next step is taking into account small irregularities in density $N$, that are inevitably present in a real plasma. Scattering by small-scale fluctuations leads to a broadening of the angular spectrum of the waves. Here we restrict ourselves to discussion of the qualitative aspects of issue. We assume that the scattering occurs at very small angles. We believe also that the average value of the scattering angle is zero. Lets denote the rms value of the scattering angle as $<\theta^2>$. Then we average $\gamma(k_\parallel, \theta)$ over $\theta$ and denote the result of averaging as $\gamma(k_\parallel)$. The value $\gamma(k_\parallel)$ is determined by the right-hand side of equation (1) in which $\theta^2$ is replaced by $<\theta^2>$. Now we expand the function $\gamma(k_\parallel)$ in the vicinity of its maximum

$$\gamma(k_\parallel) = \gamma_m \left[1 - \left(k_\parallel - k_m\right)^2 / \Delta k^2\right]. \tag{2}$$

The maximum of the increment $\gamma_m$, and the width of the instability range $\Delta k$ are determined by the parameters $\mu$, $\eta$, $k_*$, $k_0$, and $<\theta^2>$. However we will not dwell on this, since we are only interested in the overall structure of equation (2).

Qualitative picture of the evolution of small perturbations is as follows. At the linear stage small perturbations increase, the angular range is narrowed rapidly, and the wave numbers are concentrated in a small neighborhood of $k_m$. At this stage the wave is a spatially periodic structure, the IMF field lines are orthogonal to wave fronts. As the amplitude increases non-linear processes come into effect that leads to modulation instability of the wave field. The result is a broadening of the spectrum. This leads to transfer of wave energy from the amplification region ($|k_\parallel - k_m| < \Delta k$) to dissipation region ($|k_\parallel - k_m| > \Delta k$). Thus, the wave amplitude is stabilized by energy transfer over the spectrum.

Now we would like to use our theory for the formulation of an interesting prediction. We will show that the IC waves will be observed as the frequency modulated oscillation when they are detected by using the spacecraft or the ground based magnetometer.

Indeed, in the laboratory reference system the frequency equals

$$\omega = \mathbf{kU} + \omega' \qquad (3)$$

due to the Doppler effect. Since $\theta = 0$, we have $\mathbf{kU} = k_m U \cos\psi$, where $\mathbf{U}$ is the solar wind velocity, and $\psi$ is the angle between the vectors $\mathbf{U}$ and $\mathbf{B}$. In the typical cases, it is acceptable to ignore the second term on the right hand side of equation (3). Hence, we obtain the following estimate for the carrier frequency of oscillations in the laboratory system

$$\omega \approx \omega_{0p} \ U/c \ \cos\psi. \qquad (4)$$

The maximum frequency $\omega_{max} \approx (U/c)\omega_{0p}$ is reached at $\psi = 0$, when two vectors $\mathbf{U}$ and $\mathbf{B}$ are parallel to each other. For example, $\omega_{max} \sim 5\,\text{s}^{-1}$ when $U \sim 4\cdot 10^7$ cm s$^{-1}$, and $\omega_{0p} \sim 4\cdot 10^3$ s$^{-1}$. During the propagation of large-scale high-amplitude Alfvén waves from the Sun slow variation of the angle $\psi$ occurs in a wide range. This leads to profound modulation of IC wave carrier frequency. The frequency $\omega$ varies from zero to $\omega_{max} \sim 5\,\text{s}^{-1}$. So, an observer at rest sees frequency modulated emission.

We presented a specific case of Doppler effect in moving homogeneous magnetized plasma. Medium velocity is constant but wave fronts vary in their direction because waves are guided by fluctuating magnetic field lines. As a result, an observer at rest sees frequency modulated emission. Such a mechanism might be implemented in the heliosphere. The quiet solar wind can be considered as nearly homogeneous moving plasma and the IMF is frozen in it. So if we have some electromagnetic waves with wave fronts orthogonal to the IMF lines we get a state described above. The theory presented in our work predicts the waves having this property really exist. They are self-excited due to small-scale electromagnetic ion cyclotron instability of interplanetary plasma.

Evidently, IMF lines might be swayed by some lower frequency oscillations. We would like to analyze a particular case of the mentioned modulation when the ion cyclotron waves excited in the solar wind experience a modulating exposure of Alfvén waves driven by photospheric motions with a characteristic period of five minutes. The lengths and periods of Alfvén waves are much greater than the lengths and periods of ion-cyclotron waves. The result presented below show that a deep frequency modulation can be observed due to the considered effect. To test the theoretical predictions we have analyzed ground-based observations of the ULF geoelectromagnetic waves in order to find the quasi-monochromatic oscillations the carrier frequency of which varies with time within a wide range.

Based on general ideas about the configuration of the magnetosphere and the probable ways of penetration of waves from the interplanetary medium into the magnetosphere, we began the search with a careful analysis of ULF waves in polar caps. This opportunity was introduced due to the fact that in the period of the International Geophysical Year (1957–1958) V.A. Troitskaya arranged the observations of the ULF oscillations at stations located in various regions of the planet, including the Arctic and the Antarctic [Troitskaya, 1961]. In this work, Valeria Troitskaya's outstanding organizational talent was manifested.

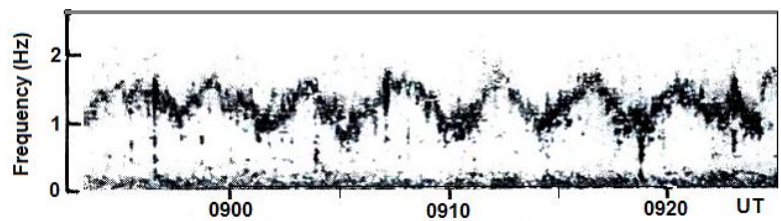

Fig. 5. Dynamic spectrum of the serpentine emission observed on 30 January 1968 at the Vostok station, Antarctica.

The so-called "serpentine emission" (SE) turns to be the most suitable type of oscillations for our purposes [Guglielmi et al., 2015a,b]. This emission is observed in the polar caps in the frequency band of 0.1–5 Hz. The SE oscillations are observed exclusively in quiet geomagnetic conditions. Clearly expressed deep modulation of the carrier frequency is a particularly important feature of the SE (Figure 5). It is not excluded that SE demonstrates a manifestation of the non-trivial Doppler effect described above.

In [Dovbnya et al., 2017], digitized data were selectively used for spectral analysis of the frequency modulation of SE oscillations. The obtained spectrum was compared with a typical spectrum of the solar surface oscillations. A similarity of these spectra was found; in both cases there is a distinct dominant 5-minute peak, which indirectly indicates the impact of solar oscillations on the ULF oscillation mode in the Earth's magnetosphere.

An intriguing question arises: did we succeed in discovering traces of oscillations of the heated surface of the Sun in the icy expanses of Antarctica? Valeria Troitskaya, we are sure, would like this idea.

**3.2. Big magnetic pulses (BMP).** So, if we accept the viewpoint set out above, then SE is an ion cyclotron wave modulated in frequency by the Alfvén wave of solar origin. We have also tried to detect a modulating wave on the Earth's surface, using the data of the ULF oscillation recording in the Pc5 range. Taking into account the fact that SE oscillations are observed under very calm geomagnetic conditions, we were looking for a rare variety of Pc5 oscillations, which would be observed under the same conditions. (Recall that typical Pc5 is observed during geomagnetic disturbances.) The preliminary search result is encouraging [Guglielmi, Potapov, 2017b]. A suitable candidate was the so-called Big Magnetic Pulses, or BMP, which sporadically occur in high latitudes on a quiet magnetic background and are isolated large-amplitude magnetic pulses.

Let us briefly outline the morphology of BMP (see [Guglielmi, Potapov, 2017b] and the literature mentioned there). Most often, pulses are observed at high latitudes around noon, under moderate geomagnetic activity (the Kp index usually does not exceed 2.7). The maximum occurrence rate of BMP is located in the interval 70°–76° of geomagnetic latitudes. The pulse duration varies from case to case in the range from half an hour to one and a half hours, and the amplitude is usually several tens of nT. The pulse is a wave packet with a characteristic filling period of 300 s. It should be noted that BMPs are classified as rare events. The typical waiting time for BMP is approximately 10 days.

Figure 6 gives an idea of the oscillogram of BMP. The oscillations are recorded at the Mirny Observatory in Antarctica ($\Phi = -76.93^o$, $\Lambda = 122.92^o$). Let's pay attention to two important circumstances for us. First, strong oscillations are observed at low geomagnetic activity, and, secondly, the period of oscillations is close to five minutes.

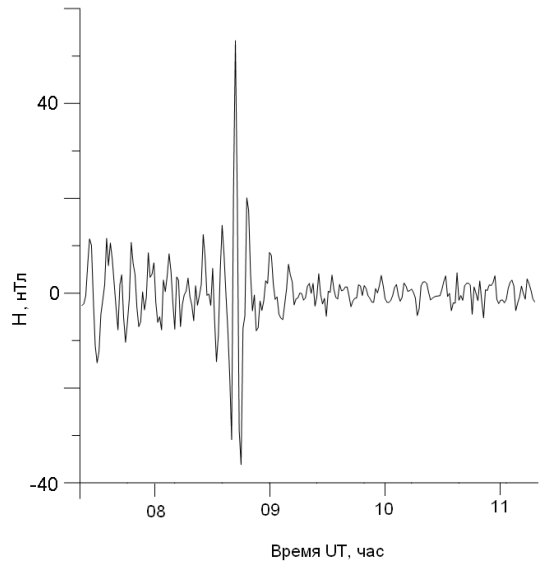

Fig. 6. ULF oscillations in the Pc5 range observed at the Mirny Antarctic Observatory on 4 January 1988 under very calm geomagnetic conditions, Kp = 1.

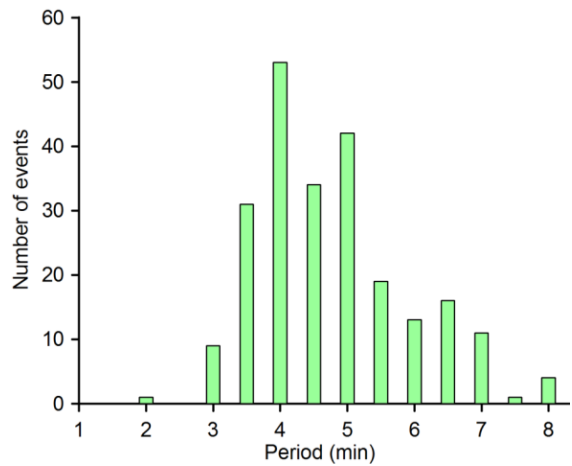

Fig. 7. Distribution of BMP by periods of oscillations calculated from observations at the obs. Mirny from 1988 to 1995.

In Borok Geophysical Observatory of IPE RAS, the richest archive of ULF oscillations is stored, which were registered in obs. Mirny during the Antarctic expeditions organized by V.A. Troitskaya [Troitskaya, 1961]. We are grateful to B.I. Klain for his kind permission to use in our work the BMP data extracted from this archive. The preliminary result of processing data accumulated from 1988 to 1995 is shown in Figure 7. We see a fairly compact distribution of BMP by period. In half of all events, the oscillation period varied from case to case in the interval from 4 to 5.4 minutes. The average value of the period is 4.8 minutes, i.e., quite close to the typical period of oscillations of the solar photosphere. So far, it is difficult to say whether we were able to detect on Earth the Alfvén waves excited on the Sun. However, there is one encouraging circumstance: Alfvén waves with a characteristic period of 4.75 min, genetically related to solar oscillations, exist in the interplanetary medium in the vicinity of the Earth [Potapov et al., 2013].

## 4. Earthquakes

So, apparently, we can say with some certainty that the 5-minute solar oscillations excite the Alfvén waves in the interplanetary medium, which act on the ULF oscillation mode of the Earth's magnetosphere. In the Introduction, with all the necessary reservations, we raised the controversial question of whether the Alfvén 5-minute waves affect the Earth's seismicity. At our request O.D. Zotov conducted a pilot analysis of global seismicity using the methodology described in [Guglielmi, Zotov, 2012a]. Surprisingly, he immediately discovered in the spectrum of earthquakes sequence a peak on the period close to the five-minute period of helioseismic oscillations!

Generally speaking, the result of the pilot analysis is not inextricably controversial with the widely known ideas of the influence of solar and geomagnetic activity on tectonic processes. In fact, more than a hundred years ago, the search for a connection between earthquakes and geoelectromagnetic variations was started [Bauer, 1906]. There is an extensive literature on this subject. We are forced to confine ourselves here only to a few publications, which, however, quite clearly reflect the interesting but difficult situation in this field of research [Tarasov et al., 2000; Hayakawa, 2001; Zakrzhevskaya, Sobolev, 2002; Duma, Ruzhin, 2003;

Fainberg et al., 2004; Hattori, 2004; Avagimov et al., 2005; Sobisevich, Sobisevich, 2010; Masci, 2011; Adushkin et al., 2012; Guglielmi, Zotov, 2012c; Schekotov et al., 2012; Buchachenko, 2014].

In the context of our problem, the study of earthquakes induced by human activity is of particular interest. We note the following observation, interesting both in itself and in connection with the above-mentioned five-minute peak in the spectrum of earthquakes activity. When analyzing the catalogs of earthquakes, a weak but statistically significant effect of synchronizing seismic activity with 15-min clock markers was observed [Guglielmi, Zotov, 2012a, b].

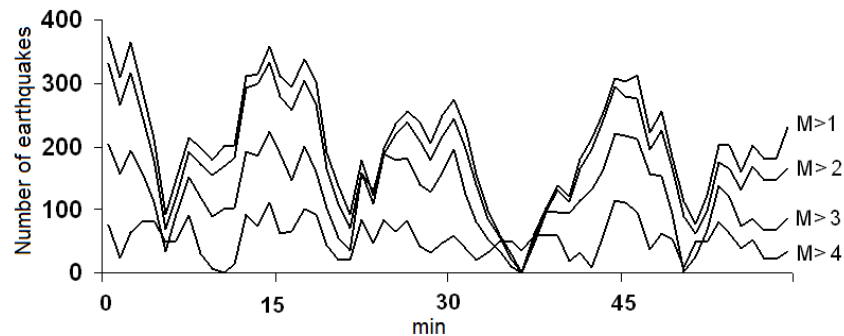

Fig. 8. 15-minute variation of seismic activity according to the USGS catalog data from 1973 to 2007.

Figure 8 explains the essence of the effect. The graph was constructed by the method of synchronous detection with an accumulation period of 60 min. To the right of each curve, the lower limit of magnitude of the earthquake M is indicated according to the USGS catalog for the period from 1973 to 2007. The horizontal axis represents the world time. Zero on the vertical scale corresponds to the numbers of earthquakes 8700, 8500, 6800 and 4600 at $M > 1, 2, 3$ and $4$, respectively. The 15-minute quasiperiodicity of weak earthquakes is clearly visible. Synchronization of seismicity with clock markers has a clearly anthropogenic origin.

Together with O.D. Zotov, we analyzed Figure 8 and other similar figures and came to the conclusion that the 5 min peak in the earthquake spectrum is only the third harmonic of the anthropogenic 15-minute seismicity variation, i.e., most likely, has nothing to do with 5-minute oscillations of the Sun. In other words, in this case we are dealing only with a numerical coincidence, which is of no interest to us. Another thing is that the synchronization of global seismicity by the world clock is interesting and incomprehensible in itself, but we will discuss this issue in the next section of our article.

## 5. Discussion

In Section 3 we briefly described the theory of frequency-modulated ion cyclotron waves in the solar wind and showed that in the polar caps of the Earth there are oscillations of SE type which are a special kind of Pc1-2 oscillations with a deep modulation of the carrier frequency. Moreover, we showed that there is a peak in the spectrum of the frequency modulation of SE at the solar oscillation frequency.

This remarkable consistency between theory and experiment can be broken if it turns out that the oscillations of SE are excited not in the interplanetary medium, but in the magnetosphere. A priori, this possibility cannot be excluded, although it is difficult to imagine how a deep modulation of the carrier frequency of the oscillations can arise in this case.

Let us leave aside the unclear question of the excitation mechanism of SE in the magnetosphere. (In the classical model, briefly presented in Section 3, the excitation mechanism is quite understandable physically.) We will concentrate on the issue of the location of SE sources. Since we are talking about oscillations with a variable frequency under very quiet magnetospheric conditions, then perhaps we can first of all suspect that the sources are located in the vents of two polar cusps. The geomagnetic field in the vents is sharply reduced and is probably variable even under quiet conditions. It is not excluded that ion cyclotron resonators are formed near the deep field minima in the cusps, similar to those described in [Guglielmi et al., 2000; Guglielmi, Potapov, 2012]. This version deserves further study, but we still do not have any productive ideas about the mechanism of excitation, nor about the mechanism of frequency modulation. The mouth of each cusp is blown by the solar wind plasma, and in this connection a distant analogy with the whistle comes to mind, but this is not enough to continue the discussion about the location of the SE sources.

Above the Pc1–2 range (T = 0.2–10 s), the Pc3 range (T = 10–45 s) is on the scale of the periods. In this range, daily permanent oscillations are observed with a carrier frequency $f$, depending on the modulus $B$ of the IMF intensity as follows:

$$f = gB. \qquad (5)$$

Here $g = 5.8 \pm 0.3$ mHz/nT [Guglielmi, 1974; Potapov, 1974] (see also the monograph [Guglielmi, Pokhotelov, 1996]). The value of $B$ is constant in the Alfven wave, so that 5-min modulation of the Pc3 carrier frequency is not expected. However, the search for hidden 5-minute amplitude modulation seems to us promising. In fact, when the Alfven wave propagates, the angle $\psi$ varies, and, generally speaking, the amplitude of Pc3 depends on it (see, for example, [Potapov et al., 2018]).

Concerning ULF oscillations in the Pc4 range (T = 45–150 s), we have neither theoretical considerations nor empirical information about the problem of interest to us. Therefore, we proceed directly to the Pc5 range (T = 150–600 s) and recall BMP oscillations, which are observed at high latitudes and have an interesting distribution over the periods (Figure 6). BMPs, like SE, are observed at low geomagnetic activity.

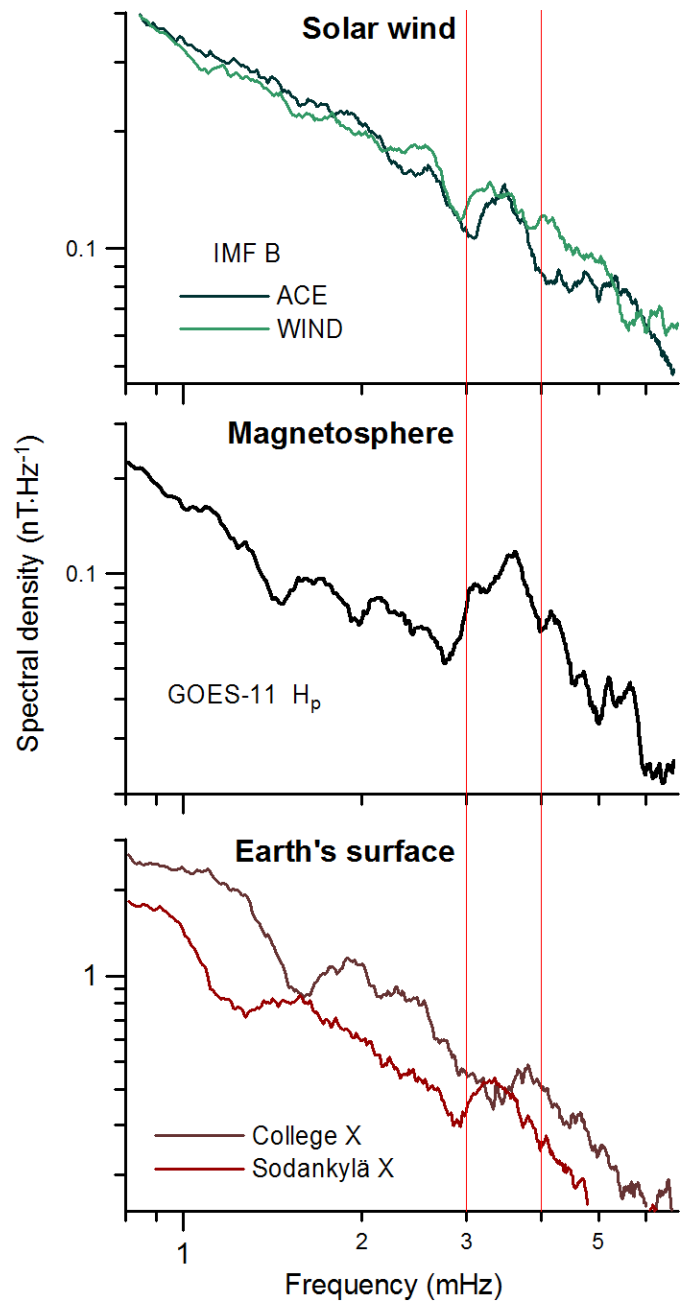

Fig. 9. Diurnal spectra of oscillations observed in three regions of space on March 13, 2009 in the Pc5 range during a weak geomagnetic storm caused by the action of a high-speed solar wind stream.

Recall that during the disturbances that occur under the influence of high-speed flows and sharp differentials in the solar wind density, the so-called global Pc5 oscillations that are observed at all geomagnetic latitudes and longitudes predominate in the magnetosphere [Potapov et al., 2006; Potapov, Polyushkina, 2010]. They are the result of excitation of the magnetospheric resonator due to the pulsed effect of the solar wind inhomogeneities. Oscillations of different periods prevail at different latitudes, and it is not possible to isolate statistically oscillations with a period of 5 minutes against this background. However, when analyzing individual events, not only the contribution of such oscillations is appreciable, but their genetic connection with Alfvén waves in interplanetary space is clearly traced. We show this using the example of Figure 9. It shows the diurnal spectra of oscillations observed in the range of 0.8–6.8 mHz on March 13, 2009 in three regions of space: in the solar wind at the Earth's orbit by two spacecraft (the upper panel), at geosynchronous orbit in the magnetosphere (middle panel) and at two high-latitude ground-based observatories (lower panel). Two vertical red lines allocate a frequency range corresponding to a period interval between 4.2 and 5.5 minutes. It can be seen that in all three media a spectral peak is clearly traced within the selected range. Moreover, in the solar wind and at geosynchronous orbit, it is the main one. On the Earth's surface, other lower-frequency peaks are evident additionally, which indicate the observation of long-period harmonics of the magnetospheric resonator at these auroral stations. Their excitation, most likely, was caused by a blow to the magnetopause of the front of a weak high-speed flow at the turn of March 12 and 13, 2009.

Now we would like to discuss earthquakes. The phenomenon of synchronism which was discovered by O.D. Zotov and summarized in Section 4 is unclear. Let us assume that the catalogs of earthquakes objectively reflect the response of tectonic activity to relatively weak but strictly periodic impulses from the technosphere synchronized according to the world time. It is known from radiophysics that under certain conditions an autooscillatory system of any physical nature can oscillate not at the frequency of auto-oscillations, but at the frequency of a weak but strictly periodic external action. In this case we say that "the frequency trapping" occurs in the oscillator. It is suggested in the paper by Guglielmi [2015] that the Earth commits a relaxation auto-oscillation. But this is just a hypothesis. We do not know if it is true. And even if it is confirmed, the question of the modulation of seismicity by periodic pulses from the technosphere will remain open.

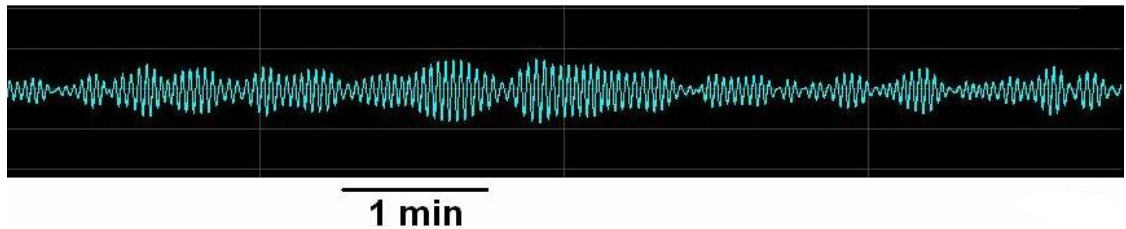

Fig. 10. Typical oscillogram of ULF Pc1 oscillations of the *pearl necklace* type.

Generally speaking, the phenomenon of synchronism was first observed when recording oscillations of the magnetosphere in the Pc1 range. It was discovered accidentally exactly 40 years ago [Guglielmi et al., 1978]. It is known that the Pc1 range extends from 0.2 Hz to 5 Hz, and so-called *pearls* are observed in it, described in detail in many reviews and monographs (see for example [Troitskaya, Guglielmi, 1967; Nishida, 1978; Guglielmi, Pokhotelov, 1996; Kangas et al., 1998]). This is one of the most beautiful phenomena of nature. The oscillogram of oscillations really reminds a string of pearls (Figure 10).

Figure 11 shows the dynamic spectra of pearls kindly prepared by B.V. Dovbnya. The figure illustrates the essence of the phenomenon of synchronism, which was studied in detail in a series of papers by the staff of the IPE RAS and the ISTP SB RAS [Guglielmi et al., 1978, 2011; Guglielmi, 1979; Zotov, Kalisher, 1979; Guglielmi, Zotov, 2010, 2012a, 2012b; Zotov, Guglielmi, 2010]. Here, events were selected, when it is clearly and undoubtedly seen that sometimes the beginning of "string of pearls", and sometimes their ending almost exactly coincides with the hour markers. Moreover, the top panel shows that the beginning of a weak series of pearls coincided with a half-hour mark, i.e. a series of oscillations began at 02:30 according to the Universal Time. In addition, there are some evidences of an increased probability of coincidence of the beginning or end of the Pc1 series with 15-minute time stamps.

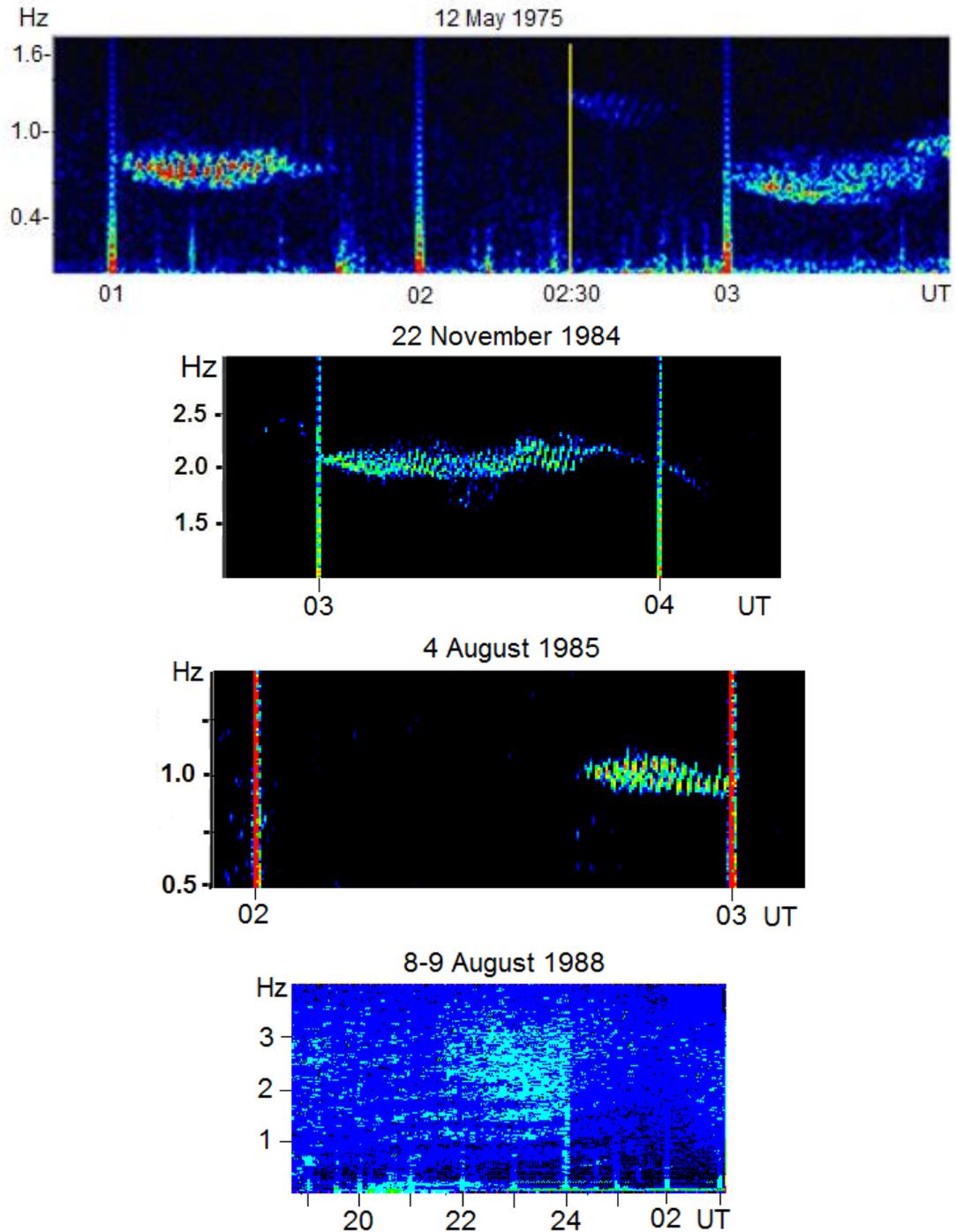

Fig. 11. Dynamic spectra of ULF Pc1 oscillations recorded at the Borok Observatory.

Looking at these spectra, it is difficult to get rid of impression that the operation of the Big Ben in London stimulates or suppresses excitation of Pc1 in the most paradoxical way. Nice idea, fantastic! The phenomenon of synchronism in the activity of Pc1 is called the Big-Ben effect. The names of the chime effect, the effect of clock marks, clock pulse effect, clock on/off effect and the like are also used, but none of them have so far become popular [Guglielmi, Pokhotelov, 1996].

The connection of the Pc1 beginnings and endings with hour marks and other moments mentioned here is rarely observed, but it is so expressive that it is difficult to believe in simple coincidences. However, the Big Ben effect not only does not correspond to the generally accepted ideas about the origin of Pc1, but it would seem to contradict even common sense.

We recall that pearls are excited in the outer radiation belt, propagate downward along the geomagnetic field lines, penetrate the ionospheric waveguide, and, propagating along the waveguide along the earth's surface, reach the observation point. And it turned out that, contrary to common sense, the mode of excitation and/or propagation of pearls is controlled by the battle of the chimes. Indeed, the easiest way would be to explain the observations by accidental coincidence. This explanation was sometimes offered to us in the corridors of conferences. But this is difficult to agree with, because the result of the analysis indicates the statistical significance of the Big Ben effect. In addition, it is useful to say that the effect, apparently, can be observed everywhere. It was noted at the mid-latitude observatories of Borok and Mondy, as well as at the high-latitude observatories Tiksi, College, Sodankylä and Vostok.

But after all this is incomprehensible! What does this means: The brink of Physics, or a verge of our ability to understand the Nature? Truly, one can exclaim "He who would search for pearls must dive below", as John Dryden said it on another occasion 340 years ago.

The difficult task is the interpretation of the Big Ben effect which is most probably due to some anthropogenic factors. We formulated the hypothesis that the beginning of each hour and other selected moments of the Universal Time serve as a kind of starting signals that globally synchronize the functioning of energy systems in such a way that sufficiently powerful electromagnetic pulses are generated that penetrate into space and, from time to time, act on the oscillatory regime of the magnetosphere. In other words, the basic idea is that the Pc1 are stimulated partly by the operation of the world network of radio transmitters and/or powerful electric devices. The beginning of every hour and other chosen moments of time serve as peculiar signals that synchronize the operation regime of the world network of the radio transmitters, in particular those that are used in radio-sounding of the ionosphere. But the physical mechanisms which lead to a corresponding modification of geophysical media due to the human activity are under a question. Thus, we consider the Big Ben effect to be a real geophysical phenomenon. The effect indicates a permanent impulse impact of the technosphere on the magnetosphere. But what exactly is a modulating factor, and as a result of which processes does the influence of this factor on the near-Earth environment lead to the appearance of Pc1 waves?

Considering the first question, only the most general judgments can be expressed so far. It is known that powerful non-coherent scattering radars entering the world network produce synchronously coordinated observations on the international program. The automatic ionospheric stations of the world network produce radio soundings of the ionosphere simultaneously every 15 minutes also according to an agreed international program.

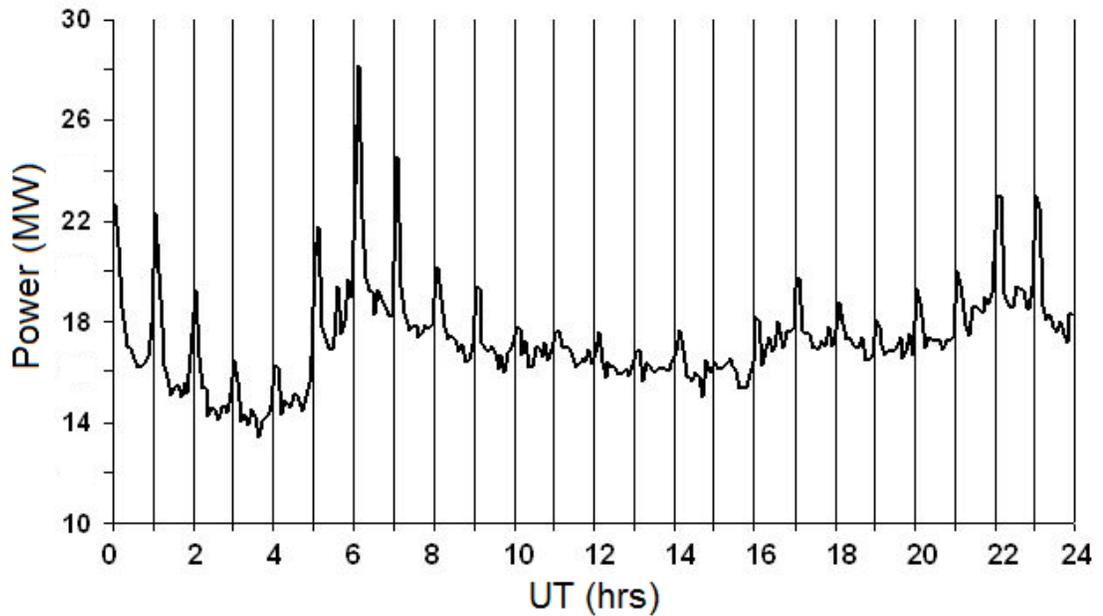

Fig. 12. Hourly pulsations of energy consumption in one of the largest industrial regions of the planet. The graph is constructed by the superposed epoch method.

Undoubtedly, there are also other factors. A special study was made by Guglielmi and Zotov [2012b], the result of which clearly demonstrates the idea of global excitation of high-power electromagnetic pulses during synchronous switching of the regime of energy systems at the zero minute of each hour. (A stronger assumption about the impulsive 15-minute modulation of the operation of the world energy system requires additional study.) In Figure 12 we see sharp peaks of the consumed electric power associated with the hour markers. It is understandable that pulsations are created spontaneously by many millions of energy consumers living by the clock and unconsciously forming the effect of chimes in the technosphere.

As for the question of the stimulation of Pc1 waves by electromagnetic pulses from the technosphere, the general idea here is that the radiation belt is in a metastable state and goes into self-excitation upon imposing a sufficiently strong pulse from texhnosphere.

So, we have no doubt that the synchronism effect in Pc1 activity indicates a noticeable impact of the technosphere on the near-earth plasma. The mechanism of such an impact is not yet known reliably. Therefore, the following energy comparisons will be quite appropriate here. According to [Kapitza, 2010], the power capacity of mankind is $\sim 1.5 \cdot 10^{20}$ erg/s. This value is comparable with the total energy flux of the solar wind through the cross section of the magnetosphere and by at least 3–4 orders of magnitude greater than the average velocity of energy inflow into the magnetospheric oscillatory systems in the entire range of ultra-low frequency electromagnetic oscillations, i.e. from millihertz to several hertz [Guglielmi, Troitskaya, 1973, p. 181]. If we now take into account that the energy stored in the near-earth plasma in the form of the Pc1 oscillations is only a very small fraction of the total energy of the ULF oscillations, then our idea will not seem too far-fetched.

When trying to clarify the issue of the origin of the Big Ben effect, we considered various hypothetical schemes of anthropogenic impact on the magnetosphere and lithosphere. Among others, the hypothesis of anthropogenic modulation of the frequency of lightning discharges was analyzed, which supposedly serve as a kind of intermediate link in the chain of cause-effect relations of interest to us. For the study, a catalog of lightning discharges registered by the LDAR of the Kennedy Space Center at Cape Canaveral in Florida, USA (http://ghrc.nsstc.nasa.gov/hydro) was used. The catalog analysis revealed some signs of anthropogenic variation in the number of discharges [Zotov, Guglielmi, 2010]. The pie chart in Figure 13 shows an unexpected result: the variation with six deep minima, coinciding with the 10-minute world time marks, is clearly visible.

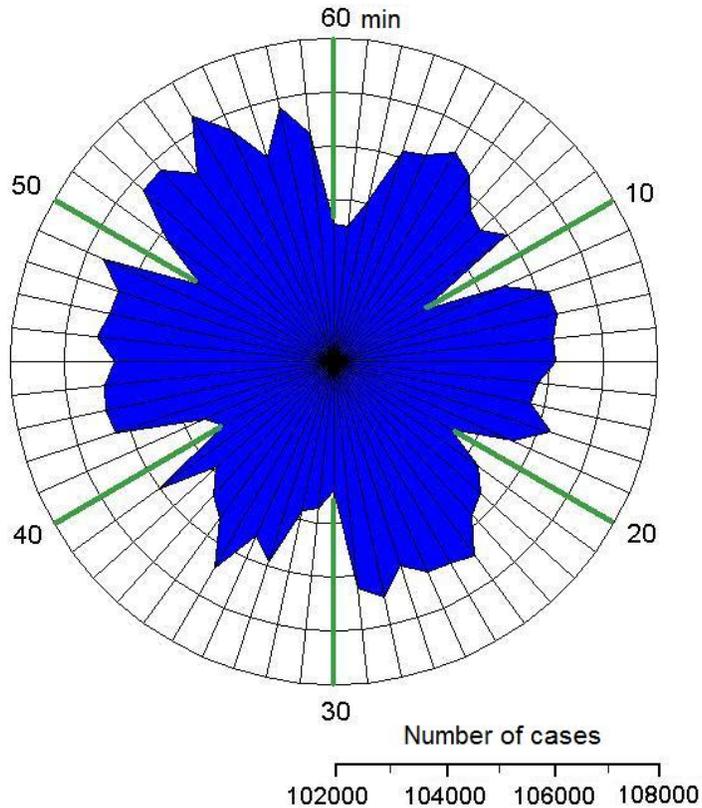

Fig. 13. Circular diagram of the occurrence rate of lightning discharges.

## 6. Conclusion

So, we give the following answer to the question we posed in the title of our paper. Apparently, the 5-minute oscillations of the Sun affect the magnetosphere, but have no effect on the lithosphere of the Earth.

If it were only a matter of this, then one could doubt the need to analyze the matter in detail. However, the analysis was useful, as it gave us an opportunity to discuss a wide range of problems relevant to the physics of solar-terrestrial relationships, the dynamics of geospheres, and, generally speaking, interesting from the general cognitive point of view.

When searching for the hidden five-minute periodicity of earthquakes, an unexpected turn of the topic arose, that led us to the problem of human impact on the environment. As a result, analysis of geophysical observations allowed us to make a plausible assumption about the existence of a hidden periodicity of humanity's energy consumption, synchronized by the world clock.

The facts described in this paper point to the need for further research. The Big Ben effect is certainly of interest from a physical point of view. In our opinion, a natural continuation of the study of the technospheric impact on the near-Earth environment would be the use of an artificial satellite to search for the effect of an increased outflow of $O^+$ ions from the ionosphere to the magnetosphere in sectors of industrial-active longitudes under the influence of ponderomotive forces [Lundin, Guglielmi, 2006]. It is expected that the upward flow of $O^+$, for example, over North America will be noticeably higher than over the Atlantic.

There is another, no less important reason that determines the relevance of further research. The point is that, in the future, control over the depth of anthropogenic modulation of natural physical processes can find application in the monitoring system of environmental degradation, giving qualitative information on long-term variation of the technogenic impact on the environment. Finally, it should be taken into account that the modulation is caused by the parasitic pulsations of energy release in the technosphere synchronized by the world clock. In this light, it can be expected that a thorough study of the anthropogenic modulation of the intensity of natural wave processes will further stimulate the development of technologies for creating energy-efficient systems for the transportation, distribution and use of energy.

At the conclusion of this paper, we would like to express judgment of a general nature. The ideas of the physics of ULF oscillations of the magnetosphere, which originated more than half a century ago, have not lost their attractiveness these days.

Moreover, they continue to develop, and the problems are enriched with new non-trivial.

*Acknowledgements*. The summer of 2018, we, the authors of this article, spent in GO Borok, located in a wonderful forest on the shore of the Rybinsk Sea. We communicated a lot with the staff of the observatory B.V. Dovbnya, B.I. Klain and O.D. Zotov, and we would like to express our deep gratitude to them for their cordial hospitality and fruitful discussions. Special gratitude is expressed to B.M. Vladimirsky, who more than 40 years ago met us in Borok and suggested the idea of investigating the effect of solar oscillations on the regime of geomagnetic variations.

The work was partially supported by Program 28 of the Presidium of the Russian Academy of Sciences, state tasks of the IPE and ISTP RAS, and by the Russian Foundation for Basic Research under the projects nos. 16-05-00056 and 16-05-00631.

# References

Adushkin V.V., Ryabova S.A., Spivak A.A., Kharlamov V.A. Response of the seismic background to geomagnetic variations. Doklady Earth Sciences. 2012. V. 444(1). P. 642–646.

Alfvén H. Cosmical electrodynamics. Oxford University Press. 1950. 237 p.

Avagimov A.A., Zeigarnik V.A., Fainberg E.B. Electromagnetically induced spatial-temporal structure of seismicity. Izvestiya. Physics of the Solid Earth. 2005. V. 41(6). P. 475–484. DOI: 10.1134/S1028334X12050157

Bauer L.A. Magnetograph records of earthquakes with special reference to the San Francisco earthquake of April 18, 1906. II. Terr. Mag. 1906. V. XI. P. 135–144.

Belcher J.W., Davis L., Jr., Smith E.J. Large-amplitude Alfvén waves in the interplanetary medium: Mariner 5. Journal of Geophysical Research. 1969. V. 74(9). P. 2302–2308.

Belcher, J. W., Davis, L., Jr. Large-amplitude Alfvén waves in the interplanetary medium, 2. Journal of Geophysical Research. 1971. V. 76(16). P. 3534–563.

Buchachenko A.L. Magnetoplasticity and the physics of earthquakes. Can a catastrophe be prevented? Physics-Uspekhi. 2014. V. 57(1) P. 92–98.
DOI: 10.3367/UFNr.0184.201401e.0101

Dovbnya B.V., Potapov A.S. Frequency modulation of serpentine emission in comparison with the set of known periods of solar oscillations. Physics of the Solid Earth. 2018. V. 5. (in press)

Dovbnya B.V., Klain B.I., Guglielmi A.V., Potapov A.S. Spectrum of frequency modulation of serpentine emission as a reflection of the solar fluctuation spectrum. Solar-Terrestrial Physics. 2017. V. **3**(1). P. 73–77.
DOI: 10.12737/article_58fd6dfaa04833.19557687

Duma G., Ruzhin Y. Diurnal changes of earthquake activity and geomagnetic Sq variation. Haz. Earth Sys. Sci. 2003. V. 3. No. 3/4. P. 171–177.

Fainberg E.B., Vasil'eva T.A., Avagimov A.A., Zeigarnik V.A. Generation of heat flows in the Earth's interior by global geomagnetic storms. Izvestiya. Physics of the Solid Earth. 2004. V. 40(4). P. 315–322.

Guglielmi A. Diagnostics of the magnetosphere and interplanetary medium by means of pulsations. Space Sci. Rev. 1974. V. 16(3). P. 331–345.

Guglielmi A.V. MHD waves in the near-earth plasma. Moscow: Nauka, 1979. 140 p.

Guglielmi A.V. On self-excited oscillations of the Earth. Izvestiya, Physics of the Solid Earth. 2015. V. 51(6). P. 920–923.

Guglielmi A.V., Pokhotelov O. A. Geoelectromagnetic Waves. Institute of Physics Publishing: Bristol (UK). 1996. 382 p.

Guglielmi A.V., Potapov A.S. The effect of heavy ions on the spectrum of oscillations of the magnetosphere . Cosmic Research. 2012. V. 50(4). P. 263–271.

Guglielmi A.V., Potapov A.S. Propagation of guided waves in moving media with application to the theory of small-scale electromagnetic waves in the solar wind plasma. In: Proceedings of "2017 Progress In Electromagnetics Research Symposium - Spring (PIERS)", St Petersburg, Russia. 2017a. P. 1051–1054.
DOI: 10.1109/PIERS.2017.8261901

Guglielmi A.V., Potapov A.S. Waves from the Sun. To the 100$^{th}$ anniversary of V.A. Troitskaya birth. Solar-Terrestrial Physics. 2017b. V. 3(3). P. 82–85.
DOI: 10.12737/stp-33201709

Guglielmi A.V., Troitskaya V.A. Geomagnetic pulsations and diagnostics of the magnetosphere. Moscow, Nauka Publ. 1973, 208 p.

Guglielmi A.V., Zotov O.D. 15-minutes modulation of the Pc1 geoelectromagnetic waves. Geomagnitnye Issledovaniya (Geophysical Research). 2010. V. 11(1). P. 64–71.

Guglielmi A.V., Zotov O.D. Spectra of hidden periodicities of the geoelectromagnetic and seismic events. Solnechno-Zemnaya Fizika (Solar-Terrestrial Physics). 2012a. Issue 20. P. 72–75.
30

Guglielmi A.V., Zotov O.D. The phenomenon of synchronism in the magnetosphere-technosphere-lithosphere dynamical system. Izvestiya, Physics of the Solid Earth. 2012b. V. 48(6). P. 486–495. DOI: 10.1134/S1069351312050035

Guglielmi A.V., Zotov O.D. Magnetic perturbations before the strong earthquakes. Izvestiya, Physics of the Solid Earth. 2012c. V. 48(2). P. 171–173. DOI: 10.1134/S1069351312010065

Guglielmi A.V., Vladimirskii B.M., Repin V.N. Geomagnetic effects of solar surface oscillations. Geomagnetizm i Aeronomiia. 1977. V. 17(5). P. 930–932.

Guglielmi A.V., Dovbnia B.V., Klain B.I., Parkhomov V.A. Stimulated excitation of Alfvén waves in the near-earth plasma by pulsed radio emission. Geomagn. Aeron. 1978. V. 18. P. 122–123.

Guglielmi A.V., Potapov A.S., Russell C.N. The ion cyclotron resonator in the magnetosphere. JETP Letters. 2000. V. 72(6). P. 298–300.

Guglielmi A.V., Dovbnya B.V., Potapov A.S., Hayakawa M. The effect of hour marks in the activity of the electromagnetic Pc1 pulsations as an evidence of anthropogenic influence on the ionosphere and magnetosphere. Solnechno-Zemnaya Fizika (Solar-Terrestrial Physics). 2011. Issue 19. P. 11–14.

Guglielmi A., Potapov A., Dovbnya B. Five-minute solar oscillations and ion-cyclotron waves in the solar wind. Solar Phys. 2015a. 290(1). P. 3023–3032. DOI 10.1007/s11207-015-0772-2

Guglielmi A.V., Potapov A.S., Dovbnya B.V. On the origin of frequency modulation of serpentine emission. Solar-Terrestrial Physics. 2015b. V. 1(2). P. 85–90. DOI: 10.12737/9617

Hattori K. ULF Geomagnetic changes associated with large earthquakes. TAO. 2004. V. 15(3). P. 329–360.

Hayakawa M. Electromagnetic phenomena associated with earthquakes: Review. Trans. Ins. Electr. Engrs. of Japan. 2001. V. 121A. P. 893–898.

Kangas J., Guglielmi A., Pokhotelov O. Morphology and physics of short-period magnetic pulsations (A Review). Space Sci. Rev. 1998. V. 83. P. 435–512.

Kapitza S.P. On the theory of global population growth. Phys. Usp. 2010. V. 53. P. 1287–1296.

Leighton R.B., Noyes R.W., Simon G.W. Velocity fields in the solar atmosphere. I. Preliminary report. Astrophys. J. 1962. V. 135. P. 474–520. DOI: 10.1086/147285

Lundin, R., Guglielmi A. Ponderomotive forces in Cosmos. Space Sci. Rev. 2006. V. 127. No. 1–4. P. 1–116. DOI: 10.1007/s11214-006-8314-8

Masci F. On the recent reaffirmation of ULF magnetic earthquakes precursors. Nat. Hazards Earth Syst. Sci. 2011. V. 11. P. 2193–2198. DOI: 10.5194/nhess-11-2193-2011

Nishida A. Geomagnetic diagnosis of the magnetosphere. New York-Heidelberg-Berlin: Springer-Verlag. 1978. 256 p.

Parker, E. N. Interplanetary dynamical processes. Interscience Publishers, New York. 1963. 272 p.

Potapov A.S. Excitation of Pc3 geomagnetic pulsations in front of the Earth's bow shock by a reflected protons beam. Issledovaniya po Geomagnetizmu, Aeronomii i Fizike Solntsa [Research on Geomagnetism, Aeronomy and Solar Physics]. 1974. Issue. 34. P. 3–12.

Potapov A.S., Polyushkina T.N. Experimental evidence for direct penetration of ULF waves from the solar wind and their possible effect on acceleration of radiation belt electrons. Geomagnetism and Aeronomy. 2010. V. 50(8). P. 950–957. DOI: 10.1134/S0016793210080049.

Potapov A., Guglielmi A., Tsegmed B., Kultima J. Global Pc5 event during 29–31 October 2003 magnetic storm. Advances in Space Research. 2006. V. 38(8). P. 1582–1586. DOI: 10.1016/j.asr.2006.05.010. ISSN 0273-1177

Potapov A.S., Polyushkina T.N., Pulyaev V.A. Observation of ULF waves on the Sun and at the Earth's orbit solar wind. Solnechno-Zemnaya Fizika (Solar-Terrestrial Physics). 2012. Issue 20. P. 45–49.

Potapov A.S., Polyushkina T.N., Pulyaev V.A. Observations of ULF waves in the solar corona and in the solar wind at the Earth's orbit. J. Atmosph. Solar-Terrestrial Phys. 2013. V. 102. P. 235–242. DOI: http://dx.doi.org/10.1016/j.jastp.2013.06.001.

Potapov A.S., Polyushkina T.N., Guglielmi A.V. Troitskaya-Bolshakova effect as a manifestation of the solar wind wave turbulence. Planetary and Space Science. 2018. V. 151. P. 78–84.

Schekotov A.Yu., Fedorov E.I., Hobara Y., Hayakawa M. ULF magnetic field depression as a possible precursor to the 2011/3.11 Japan earthquake. Telecommunications and Radio Engineering. 2012. V. 71(18). P. 1707–1718. DOI: 10.1615/TelecomRadEng.v71.i18.70

Sobisevich, L.E., Sobisevich, A.L. Dilatation structures and ULF electromagnetic quasi-oscillatory variations in the course of preparation of a strong seismic event. Vestnik ONZ RAN. 2010. V. 2. NZ6027. P. 202–213. DOI: 10.2205/2010NZ000045

Tarasov N.T., Tarasova N.V., Avagimov A.A., Zeigarnik V.A. The effect of high energy electromagnetic pulses on seismicity in Central Asia and Kazakhstan. Volc. Seis. 2000. V. 21, No. 4–5. P. 627–639.

Troitskaya V.A. Pulsations of the Earth`s electromagnetic field with periods of 1–15 sec and their connection with phenomena in the high atmosphere. J. Geophys. Res. 1961. V. 66(1). P. 5–18.

Troitskaya V.A., Guglielmi A.V. Geomagnetic micropulsations and diagnostics of the magnetosphere. Space Sci. Rev. 1967. V. 7. No. 5/6. P. 689–769.

Ulrich R.K. The five-minute oscillations on the solar surface. Astrophys. J. 1970. V. 162. P. 993–1002.

Vorontsov, S.V., Zharkov V.N. Free oscillations of the Sun and the giant planets. Sov. Phys. Usp. 1981. V. 24(8) P. 697–716. DOI: 10.1070/PU1981v024n08ABEH004837

Zakrzhevskaya N.A., Sobolev G.A. On the seismicity effect of magnetic storms. Izvestiya. Physics of the Solid Earth. 2002. V. 38(4). P. 249–261.

Zotov O.D., Guglielmi A.V. The problems of synchronism of the electromagnetic and seismic events in the magnetosphere-technosphere-lithosphere dynamic system. Solnechno-Zemnaya Fizika (Solar-Terrestrial Physics). 2010. Issue 16. P. 19–25.

Zotov O.D., Kalisher A.L. Statistical analysis of the effects of an artificial impact on the ionosphere. In: Vlijaniye Moshchnogo Radioizlucheniya na Ionosferu [Impact of the Power Radio Waves on the Ionosphere]. Apatity, KOAN USSR. Publ. 1979. P. 125.